# A Semi-Supervised Machine Learning Approach to Detecting Recurrent Metastatic Breast Cancer Cases Using Linked Cancer Registry and Electronic Medical Record Data

[1]Albee Y. Ling, [2,3]Allison W. Kurian, [2]Jennifer L. Caswell-Jin, [2]George W. Sledge Jr., [4,5]Nigam H. Shah, [4,6]Suzanne R. Tamang

[1]Stanford University, Biomedical Informatics Training Program
[2]Stanford University School of Medicine, Department of Medicine
[3]Stanford University School of Medicine, Department of Health Research and Policy
[4]Stanford University, Department of Biomedical Data Science
[5]Stanford University, Center for Biomedical Informatics Research
[6]Stanford University, Center for Population Health Sciences

Word Count: Abstract 274, Text 2,997

# Abstract

Purpose: Most cancer data sources lack information on an important outcome: metastatic recurrence. Electronic medical records (EMRs) and population-based cancer registries contain complementary information on cancer outcomes and treatment, yet are rarely used synergistically. To enable detection of metastatic breast cancer (MBC) recurrence, we applied a semi-supervised machine learning framework to linked EMR-California Cancer Registry (CCR) data.

Patients and Methods: We studied 11,459 female patients who received an incident breast cancer diagnosis from 2000-2014 and were treated at Stanford Health Care. The dataset consisted of structured data and unstructured free-text clinical notes from each patient's EMR, linked to the population-based CCR, a component of the Surveillance, Epidemiology and End Results (SEER) database. We extracted information on metastatic disease from patient notes to infer a class label and then trained a logistic regression model with regularization for metastatic recurrence classification. We evaluated model performance on an oncologist-labeled set of 146 patients.

Results: Among 11,459 patients studied, 495 (4.3%) had *de novo* stage IV MBC. Of the remaining 10,964 patients with Stage 0-III disease, 1,374 (12.5%) were classified as having recurrent MBC and 9,590 (87.5%) were classified as not having MBC. The median follow-up time is 96.3 months (mean 97.8, standard deviation 46.7). The best-performing model incorporated natural language processing of EMR-derived features with CCR-derived features and had an area under the receiver-operating characteristic curve=0.925 [95% confidence interval: 0.880-0.969], sensitivity=0.861, specificity=0.878 and overall accuracy=0.870.

Conclusion: A framework for MBC case detection combining EMR and SEER registry data achieved good sensitivity, specificity and discrimination without requiring expert-labeled examples. This approach enables population-based research on how patients die from cancer and may identify novel predictors of cancer recurrence.

# Introduction

More than 3.5 million Americans are living with breast cancer, of whom 41,070 (40,610 women and 460 men) died from the disease in 2017 (1). Despite substantial improvements in the treatment and prognosis of early-stage breast cancer, little is known about changes in survival and other outcomes of metastatic breast cancer (MBC) patients (2,3). MBC can be found at initial diagnosis (*de novo* Stage IV disease) or as a distant recurrence of the primary tumor. Electronic medical records (EMR) contain large amounts of data collected during routine medical care delivery and have the potential to generate practice-based evidence. However, it has been challenging to make use of this abundance of data in part because of difficulties in identifying which breast cancer patients have had metastatic recurrence (4). Although *de novo* MBC patients are followed by population-based cancer registries such as the California Cancer Registry (CCR) of the national Surveillance, Epidemiology and End Results (SEER) program of National Cancer Institute, *de novo* MBC represents approximately only a quarter of all metastatic patients (5,6). SEER registries are funded to record only the first few months of treatment, and continuous follow-up by any registry to assess for metastatic recurrence would be costly. Thus, there is a profound gap in our knowledge about treatment of MBC and how patients die of this disease.

Manual case detection to identify metastatic cohorts is prohibitively laborious. An informatics approach to bridge the knowledge gap would be to identify recurrent MBC patients retrospectively from administrative healthcare claims and EMRs, which contain large amounts of data collected during routine medical care delivery and have the potential to generate much needed real-world evidence. Prior studies have developed rule-based approaches that use structured data such as qualifying diagnoses, procedures and drug codes (7–11). While such approaches are simple to replicate in a new dataset, their reliability is challenged by coding bias and differential coding practices. In addition, these approaches can suffer from low sensitivity (40%-60%), despite reasonable specificity (70%-90%) (8–10). A promising alternative is emerging systems that analyze unstructured clinical text in EMR and have shown higher sensitivity and specificity (12,13). However, their limitations include a high cost of initial development, difficulty in adapting to new systems, and most significantly, the requirement for a prohibitively large amount of manually annotated training data.

We sought to develop a *semi-supervised* machine learning framework for automating MBC case detection with the potential to support population-level surveillance research across California and nationally. Semi-supervised machine learning comprises a class of supervised learning techniques that make use of unlabeled data to train machine learning models. It falls between unsupervised learning (no labeled training data) and supervised learning (completely labeled training data). It typically consists of pairing a small amount of labeled data with a large amount of unlabeled data -- or in the case of the method *distant supervision*, a distinct data source that can be used to label training examples automatically for supervised learning, in the absence of human-labeled training data.  By leveraging the complementary patient data contained in the EMR and in the California Cancer Registry, our methodological innovation extends the distant supervision paradigm described by Mintz et al. to the problem of MBC case detection (14–16).

## Methods

Data Source

The *Oncoshare* breast cancer research database comprises a three-way data linkage at the patient level. It is an integration of retrospective EMRs of Stanford Health Care (SHC) and multiple sites of the Palo Alto Medical Foundation (PAMF) in Northern California, both linked to data from California Cancer Registry, a SEER registry (17,18). Only SHC patients were included in this study. Human Subjects approval for all research reported here was obtained from the Institutional Review Boards of Stanford University and the State of California.

The structured EMR fields in Oncoshare's clinical database include each patient's diagnoses, procedures and drug orders. The unstructured EMR fields include free-text clinician notes such as medical and social histories, impressions, and visit summaries. The CCR contains detailed demographic information such as patient age, race/ethnicity, zip code and neighborhood characteristics, insurance and marital status, tumor characteristics at initial breast cancer diagnosis, and continually updated survival data which SEER obtains through linkage to the Social Security Death Master file and other national databases (17,18).

For this study, we focused on 11,459 breast cancer patients treated at SHC from 2000-2014. Descriptive information on the study population appears in Supplementary Tables S1 and S2. Supplemental Table S1. Length of follow-up in days by survival status and tumor stage at initial diagnosis in Stanford Health Care patients. Survival status was collected by California Cancer Registry as of 2014-12-31 or any later follow-up of the specific patient. Last follow-up date was the latest date of the last follow-up from California Cancer Registry, 2014-12-31 and last encounter date in Stanford Health Care' electronic medical record. A flow chart that shows how patients were analyzed by our framework appears in Figure 1. Metastatic disease that was *de novo* Stage IV was directly retrieved from the CCR. Our informatics method focused on detecting cases of metastatic recurrence and thus included only patients initially diagnosed in Stages 0-III as recorded by the CCR.

Creating an Expert-Reviewed "Gold Standard" Patient Set for Evaluation

Two board-eligible medical oncologists (A.W.K. and J.L.C.) manually reviewed de-identified EMRs from 146 female breast cancer patients to create an evaluation set: these patients' records were not used in the development of the statistical classifier. Without the knowledge of the underlying prevalence of MBC, a balanced set of patients were iteratively selected based on evidence of MBC (or lack of) from clinical notes. The oncologists determined the presence or absence of a metastatic recurrence in each patient's medical record using clinical notes, radiology reports, and pathology reports in the EMR. The most common source of information on recurrence was the most recent medical oncology or radiation oncology visit note. If there was no such note, or if this note was written more than six months before the time of chart review or the patient's death and did not indicate MBC, then more recent notes from other clinical specialties, pathology reports and imaging reports were examined. If no evidence of MBC was found after review of all these sources, then the patient was labeled as not having recurrent MBC.

Distant Supervision of MBC Classification

Our distant supervision framework exploits the Oncoshare EMR-CCR linkage. In the absence of a large number of manually annotated cases, we used one data source from the linked data, EMR, to infer a class label for metastatic recurrence. These class labels were then used to supervise the learning of a classification model using input variables from the other data source, CCR. Our methodological innovation is to extend the distant supervision paradigm described by Mintz et al., which has been applied for over a decade in the development of general domain natural language processing (NLP) and information extraction tools, to clinical case detection (15).

Step 1: Processing EMR Clinical Notes and Assigning Distant Labels

In Step 1, we used NLP-derived features to label patients that were likely to have experienced a metastatic recurrence, based on free-text patient notes in the EMR. Specifically, we adapted an open-source clinical text analysis tool, CLEVER (CL-inical EV-ent R-ecognizer), which has been validated for EMR-based information extraction tasks in prior work, to extract metastatic disease information (19). This decision was based on the efficiency of CLEVER's tagger, which facilitates the review of intermediate system output by subject matter experts and their inclusion in the development of custom clinical NLP extractors. CLEVER's source code, base terminology, and all customized components that were developed as part of this work are distributed publicly with a MIT software license on Github[1].

---

[1] https://github.com/stamang/CLEVER

Although mature clinical NLP systems exist, they can be difficult to install and must be adapted to new sources of data. Simple taggers leveraging resources such as the National Library of Medicine's Unified Medical Language System (UMLS) and SPECIALIST Lexicon tools have been shown to rival their performance and are easier to install (20). As illustrated in Supplementary Figure S4, CLEVER makes one modification to these types of general UMLS based taggers such as Noble Coder or MetaMap (21,22) in that we pre-trained word- and phrase-embedding models on clinical text to expand terminologies that are "seeded" by UMLS terms. Using language-embedding models to identify new terms that were statistically similar to the high quality UMLS seed terms, we bootstrapped the development of an enhanced terminology using an iterative and incremental development process that included two informaticists and a subject matter expert to assist in the review of candidate terms.

After our terminology for MBC information extraction was complete, we used CLEVER to annotate the corpus and extract mentions of different metastatic disease concepts that could be used to infer the presence or absence of a metastatic recurrence. We also examined their immediate contexts to determine if the target term was negated, hypothetical or an attribute of a family member and not the patient (Supplementary Figure 5). Custom classes that we developed for metastatic recurrence detection are shown in Table 2. The CLEVER rule that we developed to assign a case label to each patient was based on the positive present mention of a term from at least one of four custom word classes: "METSBONE", "METSBRAIN", "METSLIVER", "METSLUNG". These four word classes were constructed in a data-driven way from the most common sites of metastasis among our patients. In contrast to less specific word classes such as "DRECUR", which contains words and phrases that indicate a non-specific distant recurrence event, these four work classes include terms that indicate both metastatic disease and a location distant to the breast.

Step 2: Recurrent MBC Classification

In Step 2, we used the distant labels from Step 1 to train a metastatic recurrence classification model, integrating features from the CCR data. Structured fields used as input features of classification models included age, race, ethnicity, marital status, socioeconomic status, insurance type, comorbidity, year of initial breast cancer diagnosis, cancer stage, tumor grade, tumor histology, and tumor receptor status (e.g., expression of estrogen receptor (ER), progesterone receptor (PR) and human epidermal growth factor receptor 2 (HER2/neu)). Missing data in any of the structured features above were coded as a separate category. There were 427 patient-level features that were output from clinical text processing and were used as input features for patient classification. We included the total number of terms mentioned in each of the customized word classes and their frequency as positive or negative concepts, in each specific note type and across all note types in the EMR. The NLP-derived features that were used to infer the distant label were excluded from this metastatic recurrence classification model, because they would result in "learning back" our inference process.

We trained a logistic regression model with L2 regularization using glmnet package in R (23). Compared to regular logistic regression, L2 regularization smoothly shrinks regression coefficients based on regularization parameter, lambda, while retaining all input features in the model (24). Such regularization can help reduce prediction error in our case because many of our input features are likely to be correlated. The probability cutoff of the classifier was chosen to optimize the F1 score. Finally, we tested our classifier on a physician-labeled set of 146 patients (72 cases who had metastatic recurrence and 74 controls who did not have metastatic recurrence) and measured model performance using sensitivity, specificity, positive predictive value (PPV or precision), negative predictive value (NPV), and overall accuracy. In order to quantify the contribution of NLP-derived features in our classification model, we also trained an alternative classifier. Other than having different sets of input features, all aspects of the two classifiers were kept the same for fair comparisons.

## Results

Among the 11,459 patients, follow-up time ranged from 6.3 to 202.8 months with a median of 96.3 months. The mean follow-up time was 97.8 months with a standard deviation 46.7 months. There were 495 (4.3%) patients diagnosed with *de novo* Stage IV metastatic disease. In order to determine whether or not the rest of the patients had distant recurrences, we extracted additional metastatic disease information for each of the 8,892 patients who were stage 0-III at diagnosis and had at least one clinical note in the record. Among these 8,892 patients, our text mining step labeled 1,302 as having evidence of a metastatic recurrence (14.6%) and 7,590 as not having such evidence (85.4%). Using the test set of 146 manually annotated patients, our text processing step generated 15 false-positive and 8 false-negative labels, with an overall accuracy of 0.842 as shown in the first line of Table 3.

Furthermore, we trained a distant supervised classification model for metastatic recurrence using these distantly labeled patients (1,302 as MBC and 7,590 not enough evidence of MBC) using a combination of CCR and NLP-derived features. A summary of all CCR features used in our classifier is listed in Table 1. Regularization parameter lambda was chosen to be 0.0451. Using 10-fold cross validation within the training data, we obtained the highest F-1 score of 0.89 with a probability cut-off of 0.4. This cut-off was applied to be evaluated using the 146 manually annotated records in the gold-standard set. Our final classifier achieved an area under the receiver operating characteristic curve (AUC) of 0.925, with 95% confidence interval 0.880-0.969 (DeLong) (Figure 2) (25). There were 9 false-positives and 10 false-negatives, corresponding to sensitivity=0.861, specificity=0.878 and overall accuracy=0.870 (

Table 3). We benchmarked our classifier with one that includes only CCR features, keeping all other aspects of the classifier the same. We observed an improvement in all measurements as listed in third line of Table 3.

Among 11,459 patients, 1,869 (16.3%) were MBC patients and 9,590 (83.7%) were classified as non-recurrent breast cancer patients. Among 1,869 MBC patients, 495 (26.5%) were *de novo* stage IV MBC patients, while 1,374 (73.5%) were classified as recurrent MBC patients (1,302 from text processing step and 72 reviewed by physicians). This result is consistent with a recent report from the SEER registry using unrelated methods (6). Table 4 summarizes sociodemographic, clinical and genetic features of patients grouped into MBC (Stage 0-III at diagnosis), Stage IV at diagnosis, and non-MBC, based on our final metastatic recurrence classification model. The features with the highest beta coefficients from our statistical classifier are shown in Supplementary Table S3.

## Discussion

The lack of high-quality longitudinal databases that can be used to study metastatic recurrence is the biggest obstacle to practice-based evidence on how patients die from cancer. To address this problem, we developed a novel scalable framework that enables retrospective MBC case detection with good performance. The contribution of this work is three-fold. First, we retrieved information from the unstructured text of clinical notes by developing a custom NLP extraction tool for metastatic recurrence. Second, we applied a semi-supervised machine learning technique, distant supervision, to the problem of metastatic recurrence classification. Lastly, we leveraged complementary data sources, specifically medical records from Stanford Health Care and the California Cancer Registry to develop a framework for detection of metastatic recurrence that can enable population-based studies of patients with metastatic cancer.

Our classification model for recurrent MBC achieved good discriminating power and was based on a combination of features from both the CCR and EMRs. For each data source, the top predictors appear in Supplementary Table S3, by the relative rank of their model coefficient. While many of the strongest predictors from the CCR were known risk factors for metastatic recurrence (i.e., more advanced stage, triple negative disease, earlier year of diagnosis), one of the strongest classifying feature for distant recurrence was being uninsured. We found that patient insurance characteristics were essential for optimal classification of patients with metastatic disease, underscoring the importance of a diverse set of predictors and emphasizing the influence of healthcare access on cancer outcomes.

Our work suggests that an important next step is to develop tools for temporal information extraction. Due to the relatively short time between metastatic recurrence and death, NLP approaches must perform at high accuracy to support meaningful survival analysis. Although we initially planned to estimate onset time for metastatic recurrence cases, we found that simple methods (e.g., using the timestamp of a note with a positive-affirmative MBC mention) were not sufficient. Analyses of notes from ten patients found that the most common errors of this naïve approach were attributable to phrases such as "patient was diagnosed with metastatic breast cancer [number] months ago at [another medical institute]". Possible future directions for automating recurrent MBC case detection could be to acquire linguistic annotations of English clinical text or other data for training a temporal metastatic recurrence classification model.

Other limitations of this study include incomplete data due to patients receiving care outside of SHC; this is mitigated to some extent by statewide capture of treatment summaries by CCR but does not capture events outside of the state. Moreover, CCR is limited in treatment detail. Lastly, our work has primarily focused on NLP-derived features from unstructured free-text data in the EMR and structured data from the CCR. The integration of structured data from the EMR, such as diagnoses, drugs and procedures that patients received as part of their treatment and continued survivorship care, may also improve classification, especially when there is ambiguity in describing metastasis in the notes or for patients without any clinical notes (12). Also, we used a relatively simple machine learning classifier: a generalized linear model. Use of decision tree analysis and more nuanced machine learning methods may improve classification performance.

In conclusion, we developed an open-source, machine learning MBC case detection framework that accurately labels breast cancer patients as metastatic or not using linked EMR-CCR data. Our final classifier for constructing MBC cohorts leveraged both EMR and SEER data and outperformed one that uses CCR features only. As more linked datasets are developed (for example, the American Society of Clinical Oncology's CancerLinQ initiative[2]), tools such as ours can readily be adapted for them. This approach has tremendous potential to identify cohorts of metastatic cancer patients and offer insights into the characteristics, care and outcomes of this important and understudied patient population.

---

[2] https://cancerlinq.org/

# Tables

*Table 1. Characteristics of all studied breast cancer patients (N=11,459) derived from the California Cancer Registry* (26).

|  | Stage at Diagnosis | | | | | | | | | | | |
|---|---|---|---|---|---|---|---|---|---|---|---|---|
|  | Stage 0 | | Stage I | | Stage II | | Stage III | | Stage IV | | Missing | |
|  | No. | % | No. | % | No. | % | No. | % | No. | % | No. | % |
| **Total** | 2335 | 100 | 3820 | 100 | 3443 | 100 | 1120 | 100 | 495 | 100 | 246 | 100 |
| **Age at diagnosis: mean(std)** | 55.31 (11.93) | | 56.68 (12.97) | | 53.26 (13.2) | | 51.77 (12.74) | | 54.61 (13.62) | | 56.71 (15.91) | |
| *Year of breast cancer diagnosis* | | | | | | | | | | | | |
| **Before 2005** | 586 | 25.10 | 1106 | 28.95 | 1137 | 33.02 | 245 | 21.88 | 116 | 23.43 | 75 | 30.49 |
| **2005-2009** | 1039 | 44.50 | 1396 | 36.54 | 1277 | 37.09 | 474 | 42.32 | 189 | 38.18 | 118 | 47.97 |
| **2010-2015** | 710 | 30.41 | 1318 | 34.50 | 1029 | 29.89 | 401 | 35.80 | 190 | 38.38 | 53 | 21.54 |
| *Race* | | | | | | | | | | | | |
| **White** | 1762 | 75.46 | 3075 | 80.50 | 2699 | 78.39 | 871 | 77.77 | 394 | 79.60 | 191 | 77.64 |
| **Black** | 66 | 2.83 | 109 | 2.85 | 114 | 3.31 | 53 | 4.73 | 25 | 5.05 | 10 | 4.07 |
| **Asian/Pacific Islander** | 472 | 20.21 | 599 | 15.68 | 586 | 17.02 | 179 | 15.98 | 70 | 14.14 | 32 | 13.01 |
| **Other** | 21 | 0.90 | 23 | 0.60 | 17 | 0.49 | 5 | 0.45 | 2 | 0.40 | 11 | 4.47 |
| **Missing** | 14 | 0.60 | 14 | 0.37 | 27 | 0.78 | 12 | 1.07 | 4 | 0.81 | 2 | 0.81 |
| *Ethnicity* | | | | | | | | | | | | |
| **Hispanic** | 161 | 6.90 | 256 | 6.70 | 327 | 9.50 | 124 | 11.07 | 55 | 11.11 | 29 | 11.79 |
| **Non-Hispanic** | 2155 | 92.29 | 3553 | 93.01 | 3104 | 90.15 | 995 | 88.84 | 439 | 88.69 | 206 | 83.74 |
| **Missing** | 19 | 0.81 | 11 | 0.29 | 12 | 0.35 | 1 | 0.09 | 1 | 0.20 | 11 | 4.47 |
| *Neighborhood socioeconomic status (SES)** | | | | | | | | | | | | |
| **Lowest quintile** | 104 | 4.45 | 158 | 4.14 | 151 | 4.39 | 62 | 5.54 | 40 | 8.08 | 25 | 10.16 |
| Second quintile | 188 | 8.05 | 298 | 7.80 | 313 | 9.09 | 122 | 10.89 | 64 | 12.93 | 29 | 11.79 |
| Third quintile | 315 | 13.49 | 489 | 12.80 | 520 | 15.10 | 188 | 16.79 | 81 | 16.36 | 35 | 14.23 |
| Fourth quintile | 468 | 20.04 | 735 | 19.24 | 666 | 19.34 | 230 | 20.54 | 105 | 21.21 | 46 | 18.70 |
| Highest quintile | 1202 | 51.48 | 2062 | 53.98 | 1737 | 50.45 | 496 | 44.29 | 195 | 39.39 | 106 | 43.09 |
| Missing | 58 | 2.48 | 78 | 2.04 | 56 | 1.63 | 22 | 1.96 | 10 | 2.02 | 5 | 2.03 |
| *Tumor receptor subtype\*\** | | | | | | | | | | | | |
| Estrogen (ER) and/or progesterone receptor (PR)-positive and HER2-negative | 118 | 5.05 | 2336 | 61.15 | 1800 | 52.28 | 567 | 50.63 | 247 | 49.90 | 66 | 26.83 |
| HER2-positive | 48 | 2.06 | 545 | 14.27 | 651 | 18.91 | 254 | 22.68 | 117 | 23.64 | 22 | 8.94 |
| Triple-negative | 12 | 0.51 | 366 | 9.58 | 556 | 16.15 | 199 | 17.77 | 63 | 12.73 | 18 | 7.32 |
| Missing | 2157 | 92.38 | 573 | 15.00 | 436 | 12.66 | 100 | 8.93 | 68 | 13.74 | 140 | 56.91 |
| *Grade* | | | | | | | | | | | | |
| 1 | 209 | 8.95 | 1199 | 31.39 | 470 | 13.65 | 120 | 10.71 | 21 | 4.24 | 31 | 12.60 |
| 2 | 878 | 37.60 | 1551 | 40.60 | 1356 | 39.38 | 387 | 34.55 | 167 | 33.74 | 50 | 20.33 |
| 3 | 865 | 37.04 | 820 | 21.47 | 1420 | 41.24 | 521 | 46.52 | 177 | 35.76 | 64 | 26.02 |
| Missing | 383 | 16.40 | 250 | 6.54 | 197 | 5.72 | 92 | 8.21 | 130 | 26.26 | 101 | 41.06 |
| *Histology* | | | | | | | | | | | | |
| Ductal | 1989 | 85.18 | 3296 | 86.28 | 2919 | 84.78 | 923 | 82.41 | 413 | 83.43 | 168 | 68.29 |
| Lobular | 181 | 7.75 | 283 | 7.41 | 341 | 9.90 | 174 | 15.54 | 56 | 11.31 | 12 | 4.88 |
| Other | 165 | 7.07 | 241 | 6.31 | 183 | 5.32 | 23 | 2.05 | 26 | 5.25 | 66 | 26.83 |
| *Marital Status* | | | | | | | | | | | | |
| Single | 333 | 14.26 | 577 | 15.10 | 512 | 14.87 | 184 | 16.43 | 95 | 19.19 | 41 | 16.67 |
| Married | 1577 | 67.54 | 2555 | 66.88 | 2322 | 67.44 | 747 | 66.70 | 289 | 58.38 | 122 | 49.59 |
| Divorced | 184 | 7.88 | 301 | 7.88 | 293 | 8.51 | 100 | 8.93 | 47 | 9.49 | 27 | 10.98 |
| Widowed | 169 | 7.24 | 286 | 7.49 | 217 | 6.30 | 55 | 4.91 | 44 | 8.89 | 24 | 9.76 |
| Separated, Unmarried or Domestic Partner | 50 | 2.14 | 66 | 1.73 | 57 | 1.66 | 26 | 2.32 | 15 | 3.03 | 28 | 11.38 |
| Missing | 22 | 0.94 | 35 | 0.92 | 42 | 1.22 | 8 | 0.71 | 5 | 1.01 | 4 | 1.63 |
| *Payer* | | | | | | | | | | | | |
| Not insured | 11 | 0.47 | 17 | 0.45 | 22 | 0.64 | 6 | 0.54 | 3 | 0.61 | 5 | 2.03 |
| Insurance, not otherwise specified | 244 | 10.45 | 367 | 9.61 | 362 | 10.51 | 111 | 9.91 | 45 | 9.09 | 20 | 8.13 |
| Managed care/HMO/PPO | 1455 | 62.31 | 2218 | 58.06 | 2031 | 58.99 | 658 | 58.75 | 241 | 48.69 | 111 | 45.12 |
| Medicaid | 90 | 3.85 | 146 | 3.82 | 228 | 6.62 | 128 | 11.43 | 60 | 12.12 | 13 | 5.28 |
| Medicare | 424 | 18.16 | 900 | 23.56 | 639 | 18.56 | 162 | 14.46 | 118 | 23.84 | 61 | 24.80 |
| Others | 41 | 1.76 | 50 | 1.31 | 48 | 1.39 | 13 | 1.16 | 11 | 2.22 | 5 | 2.03 |
| Missing | 70 | 3.00 | 122 | 3.19 | 113 | 3.28 | 42 | 3.75 | 17 | 3.43 | 31 | 12.60 |

*\* Neighborhood socioeconomic status (SES) quintile was assigned based on a previously developed measurement by Yost et al for cases diagnosed from 2000-2005, and Shariff-Marco et al for cases diagnosed 2006-2015 (27,28).*

*\*\* Triple negative: estrogen receptor, progesterone receptor and HER2 all negative. HER2 positive: HER2 positive, regardless of estrogen receptor or progesterone receptor status.*

*Table 2. Metastatic Breast Cancer Term to Concept Mappings*

| Custom Word Class | Short Description | Example Terms |
|---|---|---|
| DRECUR | Distant recurrence | Recurrent metastatic tnbc, distant relapse, distant recurrences, distant metastatic disease involving |
| LRECUR | Local or regional recurrence | Regional recurrence, nodal recurrence, loco-regional failure, locally recur, in-breast recurrence, local recur |
| MBC | Metastatic breast cancer | Widespread metastatic breast cancer, widely metastatic triple, metastatic breast carcinoma, metastatic tnbc |
| MBCLOW | Metastatic breast cancer (low confidence) | Metastic*, metasteses*, metastatis*, metastatic lobular carcinoma, metastatic lesion, metastatic lesions, metastatic foci, mbc, met brca |
| METSBONE | Metastatic disease to the bone | Bone mets, bone metastasis, bone metasteses*, mets to spine, boney mets, diffuse skeletal mets, mets to spine |
| METSBRAIN | Metastatic disease to the brain | Metastatic disease involving the brain, brain mets, mets to brain, brain metastasis, brain metastases |
| METSLIVER | Metastatic disease to the liver | Liver mets, liver metastasis, liver metastases, hepatic mets, hepatic metastasis, hepatic metastases |
| METSLUNG | Metastatic disease to the lung | Mets to lung, pulm mets, lung mets, mbc pulm, lung metastases |
| METSNOS | Metastatic disease (distant organ not specified) | Widespread metastatic disease, stage4, newly diagnosed metastatic, stage iv |
| RECUR | Recurrence | Recur, rapid recurrence, multiple recurrences, recurrent disease, reoccurrence, reoccurring |
| DIED | Death | Passed away, expired on, deceased |

*\* these were original spellings from the clinical notes and the misspellings are left intentionally*

*Table 3. Performance of Distant Labels and Classification Model Using 146 Manually Reviewed Gold Standard Patients*

| Labels Source* | Performance Measurements** | | | | | | |
|---|---|---|---|---|---|---|---|
| | AUC | Sensitivity | Specificity | PPV | NPV | F-1 Score | Accuracy |
| **Distant Labels** | NA | 0.889 | 0.797 | 0.810 | 0.881 | 0.848 | 0.842 |
| **Classifier (CCR + EMR features)** | 0.925 | 0.861 | 0.878 | 0.873 | 0.867 | 0.867 | 0.870 |
| **Classifier (CCR features only)** | 0.789 | 0.542 | 0.824 | 0.750 | 0.649 | 0.629 | 0.685 |

*\*Distant Labels: natural language processing (NLP)-derived labels (using electronic medical records, EMR, only); Classifier (CCR + EMR features): recurrent metastatic breast cancer (MBC) classification (using NLP-derived and California Cancer Registry (CCR) features combined); Classifier (CCR features only): recurrent MBC classification (using CCR features only).*

*\*\* Note that positive predictive value (PPV), negative predictive value (NPV), F-1 score and overall accuracy are highly dependent on the prevalence of the condition, which in our case is 72/146 = 0.5. The actual prevalence of recurrent metastatic breast cancer in our study population is likely to be much lower. However, sensitivity, specificity, and area under the curve (AUC) are intrinsic properties of classifier and are insensitive to prevalence of cases* (29,30)*.*

*Table 4. Case Detection Results by Metastatic Breast Cancer Status*

|  | Recurrent MBC (Stage 0-III at diagnosis) | | Stage IV at diagnosis | | Non-MBC | |
|---|---|---|---|---|---|---|
|  | No. | % | No. | % | No. | % |
| **Total** | 1302 | 100 | 495 | 100 | 7590 | 100 |
| **Age at diagnosis: mean(std)** | 52.99 (13.04) | | 54.61 (13.62) | | 55.36 (12.99) | |
| *Year of breast cancer diagnosis* | | | | | | |
| **Before 2005** | 526 | 40.40 | 116 | 23.43 | 1738 | 22.90 |
| **2005-2009** | 463 | 35.56 | 189 | 38.18 | 2787 | 36.72 |
| **2010-2015** | 313 | 24.04 | 190 | 38.38 | 3065 | 40.38 |
| *Race* | | | | | | |
| **White** | 1025 | 78.73 | 394 | 79.60 | 5887 | 77.56 |
| **Black** | 51 | 3.92 | 25 | 5.05 | 229 | 3.02 |
| **Asian/Pacific Islander** | 206 | 15.82 | 70 | 14.14 | 1371 | 18.06 |
| **Other** | 4 | 0.31 | 2 | 0.40 | 57 | 0.75 |
| **Missing** | 16 | 1.23 | 4 | 0.81 | 46 | 0.61 |
| *Ethnicity* | | | | | | |
| **Hispanic** | 129 | 9.91 | 55 | 11.11 | 588 | 7.75 |
| **Non-Hispanic** | 1170 | 89.86 | 439 | 88.69 | 6968 | 91.81 |
| **Missing** | 3 | 0.23 | 1 | 0.20 | 34 | 0.45 |
| *Neighborhood socioeconomic status\** | | | | | | |
| **Lowest quintile** | 45 | 3.46 | 40 | 8.08 | 342 | 4.51 |
| **Second quintile** | 121 | 9.29 | 64 | 12.93 | 631 | 8.31 |
| **Third quintile** | 215 | 16.51 | 81 | 16.36 | 988 | 13.02 |
| **Fourth quintile** | 254 | 19.51 | 105 | 21.21 | 1433 | 18.88 |
| **Highest quintile** | 646 | 49.62 | 195 | 39.39 | 4015 | 52.90 |
| **Missing** | 21 | 1.61 | 10 | 2.02 | 181 | 2.39 |
| *Stage* | | | | | | |
| **0** | 72 | 5.53 | 0 | 0.00 | 1813 | 23.87 |
| **I** | 302 | 23.20 | 0 | 0.00 | 2837 | 37.37 |
| **II** | 585 | 44.93 | 0 | 0.00 | 2186 | 28.80 |
| **III** | 307 | 23.58 | 0 | 0.00 | 616 | 8.10 |
| **IV** | 0 | 0.00 | 495 | 100.00 | 0 | 0.00 |
| **Missing** | 36 | 2.76 | 0 | 0.00 | 141 | 1.86 |
| *Tumor receptor subtype*** | | | | | | |
| **Estrogen (ER) and/or progesterone receptor (PR)-positive and HER2-negative** | 608 | 46.70 | 247 | 49.90 | 3514 | 46.30 |
| **HER2-positive** | 259 | 19.89 | 117 | 23.64 | 969 | 12.77 |
| **Triple-negative** | 223 | 17.13 | 63 | 12.73 | 689 | 9.08 |
| **Missing** | 212 | 16.28 | 68 | 13.74 | 2418 | 31.86 |
| *Grade* | | | | | | |
| **1** | 163 | 12.52 | 21 | 4.24 | 1511 | 19.91 |
| **2** | 446 | 34.26 | 167 | 33.74 | 3020 | 39.79 |
| **3** | 580 | 44.55 | 177 | 35.76 | 2400 | 31.62 |
| **Missing** | 113 | 8.68 | 130 | 26.26 | 659 | 8.68 |
| *Histology* | | | | | | |
| **Ductal** | 1146 | 88.02 | 413 | 83.43 | 6433 | 84.76 |
| **Lobular** | 117 | 8.99 | 56 | 11.31 | 703 | 9.26 |
| **Other** | 39 | 3.00 | 26 | 5.25 | 454 | 5.98 |
| *Marital Status* | | | | | | |
| **Single** | 202 | 15.52 | 95 | 19.19 | 1133 | 14.93 |
| **Married** | 865 | 66.44 | 289 | 58.38 | 5096 | 67.14 |
| **Divorced** | 120 | 9.22 | 47 | 9.49 | 602 | 7.93 |
| **Widowed** | 81 | 6.22 | 44 | 8.89 | 518 | 6.82 |
| Separated, Unmarried or Domestic Partner | 20 | 1.54 | 15 | 3.03 | 170 | 2.24 |
| Missing | 14 | 1.07 | 5 | 1.01 | 71 | 0.94 |
| *Payer* | | | | | | |
| Not insured | 11 | 0.84 | 3 | 0.61 | 35 | 0.46 |
| Insurance, not otherwise specified | 144 | 11.06 | 45 | 9.09 | 729 | 9.60 |
| Managed care/HMO/PPO | 695 | 53.38 | 241 | 48.69 | 4489 | 59.14 |
| Medicaid | 121 | 9.29 | 60 | 12.12 | 390 | 5.14 |
| Medicare | 267 | 20.51 | 118 | 23.84 | 1610 | 21.21 |
| Others | 21 | 1.61 | 11 | 2.22 | 115 | 1.52 |
| Missing | 43 | 3.30 | 17 | 3.43 | 222 | 2.92 |

*\* Neighborhood socioeconomic status (SES) quintile was assigned based on a previously developed measurement by Yost et al for cases diagnosed from 2000-2005, and Shariff-Marco et al for cases diagnosed 2006-2015 (27,28).*

*\*\* Triple negative: estrogen receptor, progesterone receptor and HER2 all negative. HER2 positive: HER2 positive, regardless of estrogen receptor or progesterone receptor status.*

Figures

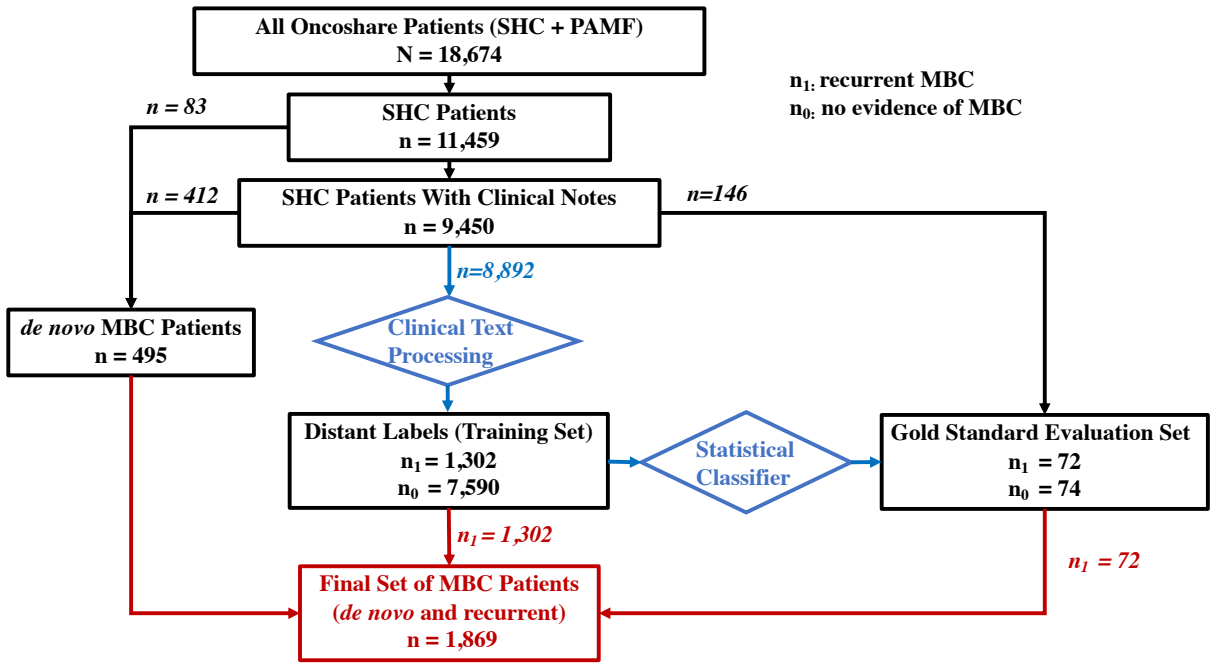

Figure 1. Flowchart of Oncoshare Patient Count by Step. SHC: Stanford Health Care; PAMF: Palo Alto Medical Foundation; MBC; metastatic breast cancer.

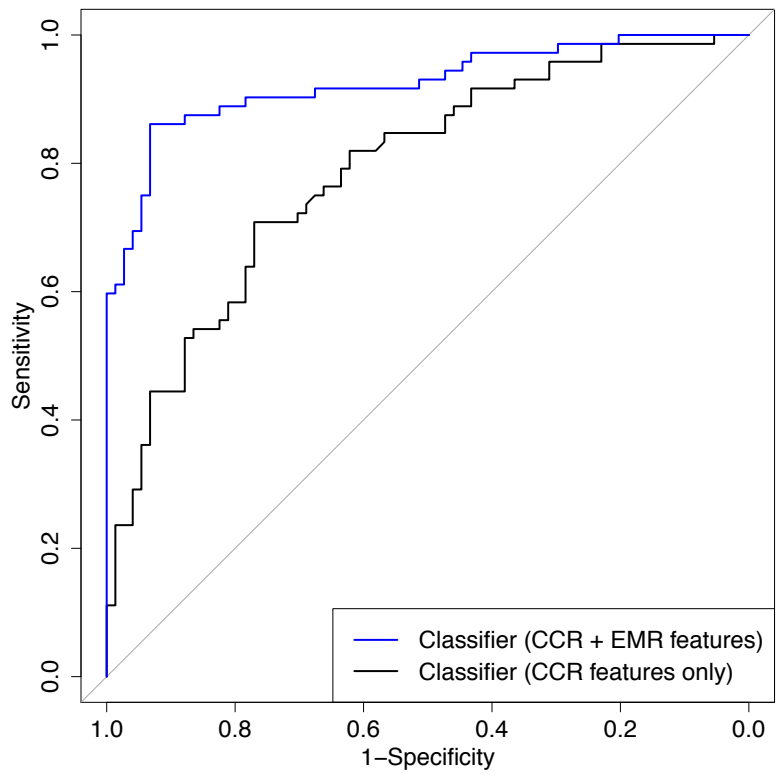

*Figure 2. Receiver Operating Characteristic Curve (ROC) of Statistical Classifiers Evaluated using the Test Set of 146 Patients. The area under the curve (AUC) of classifier with CCR and EMR features is 0.925 with 95% confidence interval 0.880-0.969. The AUC of the classifier with CCR features only is 0.789 with 95% confidence interval 0.716-0.861.*

## Acknowledgements

The authors gratefully acknowledge research support from the Breast Cancer Research Foundation; the Suzanne Pride Bryan Fund for Breast Cancer Research; the BRCA Foundation; the Jan Weimer Junior Faculty Chair in Breast Oncology; the Susan and Richard Levy Gift Fund; the Regents of the University of California's California Breast Cancer Research Program (16OB-0149 and 19IB-0124); and the National Cancer Institute's Surveillance, Epidemiology and End Results Program under contract HHSN261201000140C awarded to the Cancer Prevention Institute of California. The collection of cancer incidence data used in this study was supported by the California Department of Health Services as part of the statewide cancer reporting program mandated by California Health and Safety Code Section 103885; the National Cancer Institute's Surveillance, Epidemiology, and End Results Program under contract HHSN261201000140C awarded to the Cancer Prevention Institute of California, contract HHSN261201000035C awarded to the University of Southern California, and contract HHSN261201000034C awarded to the Public Health Institute; and the Centers for Disease Control and Prevention's National Program of Cancer Registries, under agreement #1U58 DP000807-01 awarded to the Public Health Institute. JLC was supported by an ASCO Young Investigator Award from the Conquer Cancer Foundation. AYL acknowledges Stanford Graduate Fellowship. The ideas and opinions expressed herein are those of the authors, and endorsement by the University or State of California, the California Department of Health Services, the National Cancer Institute, or the Centers for Disease Control and Prevention or their contractors and subcontractors is not intended nor should be inferred.

# Appendix

*Supplemental Table S1. Length of follow-up in days by survival status and tumor stage at initial diagnosis in Stanford Health Care patients. Survival status was collected by California Cancer Registry as of 2014-12-31 or any later follow-up of the specific patient. Last follow-up date was the latest date of the last follow-up from California Cancer Registry, 2014-12-31 and last encounter date in Stanford Health Care' electronic medical record.*

|         |      | Stage 0 | Stage 1 | Stage 2 | Stage 3 | Stage 4 | Missing | Total |
|---------|------|---------|---------|---------|---------|---------|---------|-------|
| **Alive**   | *n*    | 2162    | 3351    | 2668    | 695     | 129     | 172     | 9177  |
|         | min  | 243     | 190     | 282     | 401     | 241     | 393     | 190   |
|         | max  | 5932    | 6053    | 6083    | 5778    | 5614    | 5921    | 6083  |
|         | mean | 2893    | 2847    | 2887    | 2547    | 2200    | 2957    | 2840  |
|         | sd   | 1343.06 | 1432.91 | 1421.03 | 1268.84 | 1258.82 | 1368.73 | 1397.83 |
| **Dead**    | *n*    | 168     | 468     | 771     | 423     | 360     | 74      | 2264  |
|         | min  | 562     | 563     | 261     | 459     | 348     | 502     | 261   |
|         | max  | 5684    | 5748    | 5692    | 5466    | 5468    | 5470    | 5748  |
|         | mean | 3530    | 3543    | 3631    | 2928    | 2770    | 3414    | 3330  |
|         | sd   | 1247.06 | 1289.64 | 1268.87 | 1273.46 | 1391.04 | 1224.44 | 1336.85 |
| **Missing** | *n*    | 5       | 1       | 4       | 2       | 6       | 0       | 18    |
|         | min  | 1094    | 4971    | 523     | 464     | 560     | NA      | 464.0 |
|         | max  | 4182    | 4971    | 5222    | 1379    | 3973    | NA      | 5222.0 |
|         | mean | 2577    | 4971    | 2351    | 921.5   | 1356    | NA      | 2069.0 |
|         | sd   | 1433.99 | NA      | 2043.13 | 647.00  | 1308.64 | NA      | 1634.97 |
| **Total**   | *n*    | 2335    | 3820    | 3443    | 1120    | 495     | 246     | 11459 |
|         | min  | 243     | 190     | 261     | 401     | 241     | 393     | 190   |
|         | max  | 5932    | 6053    | 6083    | 5778    | 5614    | 5921    | 6083  |
|         | mean | 2938    | 2933    | 3053    | 2688    | 2604    | 3094    | 2935  |
|         | sd   | 1346.22 | 1434.45 | 1423.14 | 1284.59 | 1384.07 | 1341.04 | 1400.37 |

*Supplemental Table S2. Length of Follow-up in days by Survival Status and Insurance Status in Stanford Health Care patients*

|  |  | Insured* | Not Insured | Others/Missing | Total |
|---|---|---|---|---|---|
| **Alive** | *n* | 8691 | 40 | 446 | 9177 |
|  | **min** | 190 | 641 | 365 | 190 |
|  | **max** | 6083 | 5622 | 5978 | 6083 |
|  | **mean** | 2815 | 2935 | 3301 | 2840 |
|  | **sd** | 1393.37 | 1376.43 | 1409.20 | 1397.83 |
| **Dead** | *n* | 2125 | 24 | 115 | 2264 |
|  | **min** | 261 | 1456 | 714 | 261 |
|  | **max** | 5748 | 5403 | 5464 | 5748 |
|  | **mean** | 3313 | 3863 | 3534 | 3330 |
|  | **sd** | 1337.65 | 1136.35 | 1330.85 | 1336.85 |
| **Missing** | *n* | 16 | 0 | 2 | 18 |
|  | **min** | 464 | NA | 778 | 464.0 |
|  | **max** | 5222 | NA | 1094 | 5222.0 |
|  | **mean** | 2210 | NA | 936 | 2069.0 |
|  | **sd** | 1683.36 | NA | 223.45 | 1634.97 |
| **Total** | *n* | 10832 | 64 | 563 | 11459 |
|  | **min** | 190 | 641 | 365 | 190 |
|  | **max** | 6083 | 5622 | 5978 | 6083 |
|  | **mean** | 2912 | 3283 | 3340 | 2935 |
|  | **sd** | 1397.23 | 1359.90 | 1400.46 | 1400.37 |

*Insured: Insurance, NOS, Managed care/HMO/PPO, Medicaid, and Medicare

*Supplemental Table 3. Top Classification Features by Data Source and Relative Rank*

| Rank | Source | Data Field | Regression Coefficient | Exponential of Regression Coefficient |
|------|--------|------------|------------------------|----------------------------------------|
| 1 | Electronic Medical Record (EMR) Clinical Notes | MBCuniPN (Number of unique mentions of terms in MBC class) | 3.211 | 24.797 |
| 2 | EMR Clinical Notes | METSNOSPosCyto (Number of positive metastatic mentions in cytology notes) | 0.994 | 2.701 |
| 3 | EMR Clinical Notes | MBCPosOutSR (Number of positive MBC mentions in outpatient screen review notes) | 0.818 | 2.266 |
| 4 | EMR Clinical Notes | LRECURPosOutSR (Number of positive loco-regional recurrence mentions in outpatient screen review notes) | 0.746 | 2.109 |
| 5 | EMR Clinical Notes | LRECURPosMNR (Number of positive loco-regional recurrence mentions in magnetic nuclear resonance notes) | 0.677 | 1.968 |
| 11* | California Cancer Registry (CCR) | Stage 3 (reference level Stage 0) | 0.379 | 1.461 |
| 24 | CCR | Payer Not Insured (reference level missing insurance status) | 0.177 | 1.194 |
| 28 | CCR | Stage 2 (reference level Stage 0) | 0.153 | 1.166 |
| 29 | CCR | Diagnosis Before 2005 (reference level diagnosis year 2005-2009) | 0.149 | 1.161 |
| 40 | CCR | Triple Negative (reference level Estrogen (ER) and/or progesterone receptor (PR)-positive and HER2-negative) | 0.119 | 1.126 |
| 42 | CCR | Socioeconomic status 3rd quintile (reference level 1st quintile) | 0.113 | 1.119 |

\* For example, the odds of having recurrent metastatic breast cancer (MBC) for patients with Stage 3 at diagnosis are 46.1% higher than that of Stage 0, holding all other features constant.

\*\* Since mentions of words from these four word classes "METSBONE", "METSBRAIN", "METSLIVER", and "METSLUNG" were used to determine the distand labels, none of the input features into the classifier used information regarding to features in these four word classes.

\*\*\* Compared to logistic regression models without regularization, interpretation of these coefficients needs to be proceeded with caution, as L2 regularization reduces the regression coefficients of correlated variables.

*Supplemental Figure S4. Corpus-driven Expansion of Metastatic Breast Cancer Seed Terms Using Word and Phrase Embeddings*

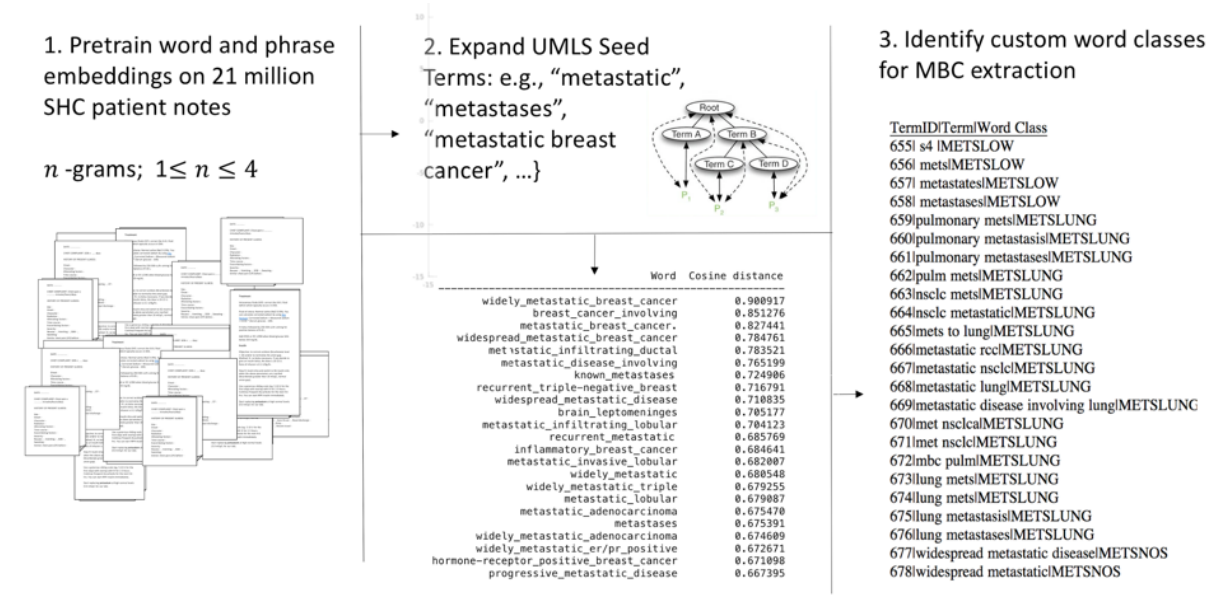

*Supplemental Figure S5. Example Text Snippet Processed by CLEVER To Extract Information On Metastatic Disease*

**SNIPPET EXAMPLE:**
"spect/ct of the pelvis and lower back was performed. findings: there are no foci of radiotracer uptake concerning for skeletal metastases."

**CLASS SEQUENCE (target class, section, other CLEVER classes and character offsets)**
NOTEID   METSBONE:1186   755   findings:690   DOT:0:687:-68   PUNCT:3:698:-57
   NEGEX:516:700:-55   HYP:576:740:-15   DOT:0:774:19

**TARGET CLASS:** METSBONE
**TARGET TERM:** skeletal metastases
**WORD CLASS SEQ:** DOT_PUNCT_NEGEX_HYP_#METSBONE#_DOT
**TAGGED TEXT:** period:DOT:0:687:-68, colon:PUNCT:3:698:-57, there are no:NEGEX:516:700:-55, concerning for:HYP:576:740:-15, period:DOT:0:774:19